# Orbital-engineered $p_{x,y}$-kagome lattice in a halogen monolayer

Xulin Liu,[1,2,#] Jingyi Duan,[1,2,#] Yueqian Chen,[1,2,#] Wenbo Liu,[3,4] Peiyao Xiao,[1,2] Yuxiang Liu,[1,2] Pei Liu,[1,2] Minjun Wang,[1,2] Baojie Feng,[3,4] Dongfei Wang,[1,2,§] Xun Shi,[1,2,*] Wei Jiang,[1,2,5,‡] Yugui Yao,[1,2,5] Wende Xiao,[1,2,†]

[1]Key Laboratory of Advanced Optoelectronic Quantum Architecture and Measurement, Ministry of Education, School of Physics, Beijing Institute of Technology, Beijing 100081, China
[2]Beijing Key Laboratory of Quantum Matter State Control and Ultra-Precision Measurement Technology, Beijing Institute of Technology, Beijing 100081, China
[3]Institute of Physics, Chinese Academy of Sciences, Beijing 100081, China
[4]School of Physical Sciences, University of Chinese Academy of Sciences, Beijing 100049, China
[5]International Center for Quantum Materials, Beijing Institute of Technology, Zhuhai 519000, China

[#]These authors contributed equally to this work.
[§]contact author: dfwang@bit.edu.cn
[*]contact author: shixun@bit.edu.cn
[‡]contact author: wjiang@bit.edu.cn
[†]contact author: wdxiao@bit.edu.cn

**ABSTRACT**. Multi-orbital kagome lattices with explicit orbital degrees of freedom remain largely unexplored, as most experimentally realized systems rely on complex *d*-electron manifolds that are approximated by isotropic single-orbital models. Here, we overcome this limitation by realizing a $p_{x,y}$-orbital kagome lattice through deposition of a Br monolayer on Ag(111), where orbital filtering selectively suppresses the $p_z$ channel. Scanning tunneling microscopy, angle-resolved photoemission spectroscopy, and density-functional-theory calculations reveal a large-area, highly ordered kagome structure whose band dispersions quantitatively match the anisotropic $p_{x,y}$ tight-binding model. To extract the intrinsic manifold from the substrate background, we construct an effective H-passivated model, which uncover the intrinsic electronic structure and reveals nontrivial topological characteristics of the $p_{x,y}$ kagome manifold driven by first-order spin-orbit coupling effect. Our work establishes Br/Ag(111) as an experimentally accessible platform for multi-orbital kagome physics, extending the kagome paradigm from the conventional *d*-orbital regime to an orbitally engineered topological setting.

## I. INTRODUCTION.

Kagome materials have emerged as a fertile platform for exotic quantum phases arising from the interplay of lattice geometry, topology, and electronic correlations [1–6]. To date, the vast majority of experimentally established transition-metal based kagome compounds, including the $AV_3Sb_5$ family [7–13], $RMn_6Sn_6$ [14–17] and $Co_3Sn_2S_2$ [18–23], are predominantly rooted in *d* orbitals [Fig. 1(a)]. In these systems, the low-energy physics are notoriously convoluted by multiple orbital channels [24,25], strong covalent hybridization with ligands [26], significant interlayer coupling [27], and pronounced spin-orbit coupling [28–30]. While these *d*-electron platforms have yielded remarkable discoveries [31–34], their inherent complexity often obscures the direct correspondence with the idealized, isotropic single-orbital kagome model [Fig. 1(b)]. More critically, they complicate the systematic exploration of genuine multi-orbital kagome physics, where the orbital degree of freedom itself becomes an active player rather than a passive perturbation [35–39].

Towards removing the strong correlated effect for *d*-electron heavy metals, kagome lattice with weak correlation *p*-orbital appears as a promising material platform. Moreover, unlike their *d*-orbital counterparts, which suffer from strong exchange splitting and crystal-field multiplicity, directional *p*-orbitals are more spatially anisotropic [40–42], which can be selectively engineered through atomic adsorption on crystalline substrates [43–45]. This "orbital-filtering" approach enables a correlation-free, orbital-active kagome lattice governed primarily by geometry and orbital phase, providing a testbed for multi-orbital kagome models [46]. However, a persistent challenge has been the scalable synthesis of such *p*-orbital kagome lattice with sufficiently long-range order to enable momentum-resolved spectroscopic characterization.

Here, we address this challenge by realizing a Br-$p_{x,y}$ kagome platform on Ag(111). Using scanning tunneling microscopy/spectroscopy (STM/STS), we confirm the formation of a large-area and highly ordered kagome monolayer with a 3×3 superstructure relative to the substrate. Angle-resolved photoemission spectroscopy (ARPES) directly resolves the characteristic $p_{x,y}$-derived kagome bands, whose dispersions deviate markedly from the isotropic single-orbital model and instead exhibit the anisotropic hopping signatures expected for a two *p* orbitals system. To disentangle the intrinsic Br kagome physics from the Ag background, we construct an effective H-passivated Br kagome model that faithfully reproduces the experimental band structure. Within this framework, the unique first-order spin-orbit coupling (SOC) opens multiple sizable global gaps with nontrivial $\mathbb{Z}_2$ indices, giving rise to robust in-gap edge states and a pronounced intrinsic spin Hall response. Our results establish Br/Ag(111) as a clean, experimentally accessible multi-orbital kagome platform that directly complements the extensively studied *d*-orbital kagome compounds and opens a new avenue toward orbital-engineered topology.

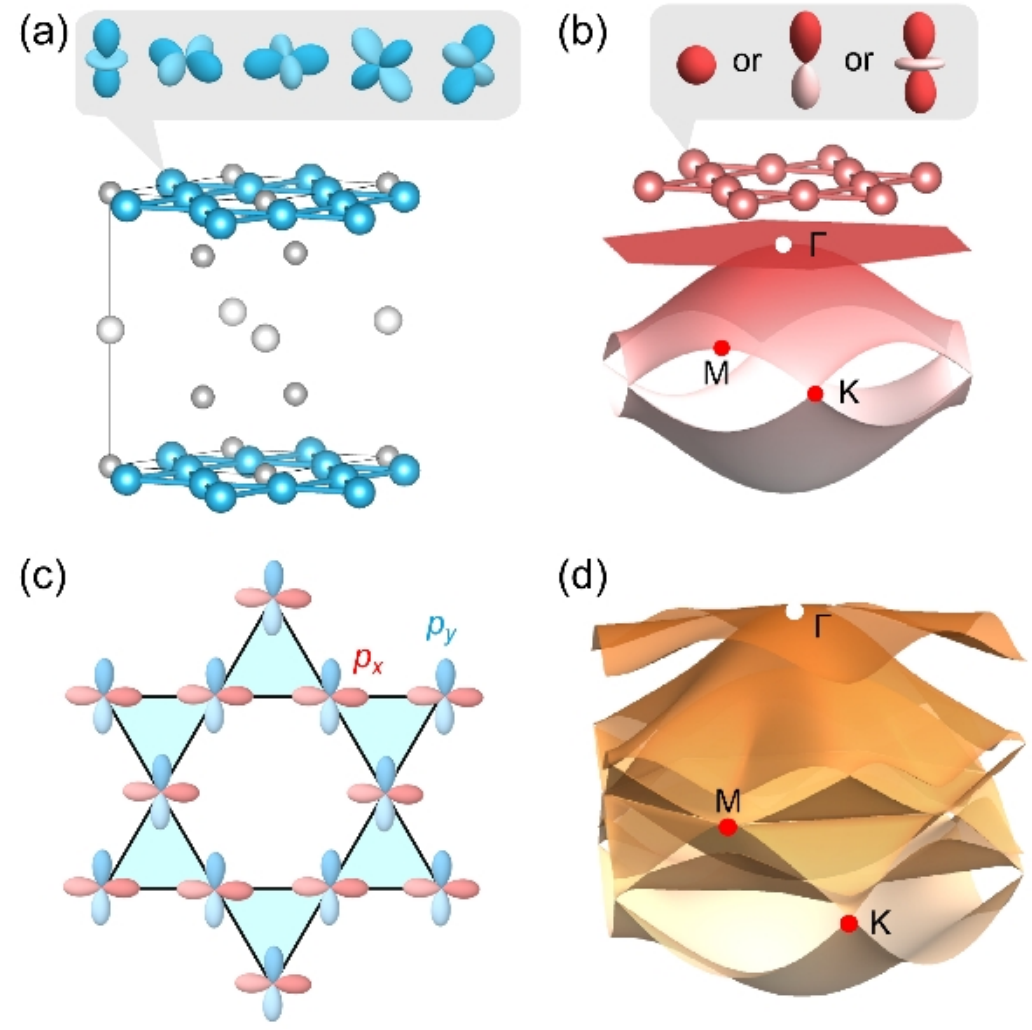

FIG. 1. Conventional kagome and $p_{x,y}$-kagome models. (a) Schematic illustration of a transition-metal *d*-orbital kagome system. (b) Three-dimensional (3D) band structure of the isotropic single-orbital ($s/p_z/d_{z^2}$) kagome model typically adopted to describe the kagome physics in (a). (c) and (d) Schematic illustration of a $p_{x,y}$-orbital kagome system and its 3D band structure.

## II. RESULTS

### A. Theoretical Model

To capture the orbital-active physics of our system, we construct a minimal tight-binding model for the in-plane $p_x$ and $p_y$ orbitals on the kagome lattice [Fig. 1(c)]. The effective Hamiltonian with only the nearest-neighbor (NN) hopping term is describe as

$$\mathrm{H}=\sum_{i}\sum_{a}\sum_{s}\epsilon_a\, c^{\dagger}_{i,p_a,s}c_{i,p_a,s}+$$

$$\sum_{<i,j>}\sum_{a,b}\sum_{s}(t_{ia,jb}\, c^{\dagger}_{i,p_a,s}c_{j,p_b,s}+h.c),$$

where $i$ and $j$ label kagome lattice sites, $a,b$=$x,y$ denotes the two in-plane orbitals, $s$=↑,↓ is the spin index, and $\epsilon_a$ denotes the on-site energy. The hopping matrix element $t_{ia,jb}$ incorporates both intra- and inter-orbital NN hopping processes. These matrix elements are parameterized by the Slater-Koster integrals $t_{pp\sigma}$ and $t_{pp\pi}$, as detailed in the model part of Supplemental Material (SM) [47]. Unlike the isotropic single-orbital model, this formulation naturally yields hopping amplitudes that vary strongly with bond orientation, directly reflecting the directional character of the $p$-orbital lobes. For a hopping-parameter ratio of $t_{pp\pi}/t_{pp\sigma}$=−0.15, the band structure shown in Fig. 1(d) is obtained.

The on-site SOC projected onto the $p_{x,y}$ subspace is

$$H_{\mathrm{SOC}}=\lambda\sum_{i}\left(ic^{\dagger}_{i,p_x,\downarrow}c_{i,p_y,\downarrow}-ic^{\dagger}_{i,p_x,\uparrow}c_{i,p_y,\uparrow}\right)+h.c.,$$

which corresponds to the local $L_zS_z$ coupling within the $p_{x,y}$ orbital subspace with $\lambda$ denoting the effective SOC strength. The total Hamiltonian is therefore $\mathcal{H}$=$H$+$H_{\mathrm{SOC}}$, which provides a direct framework for analyzing the measured band dispersions. As we show below, the SOC lifts the Dirac degeneracies inherent to the kagome geometry and drives the system into a topological nontrivial phase, characterized by a nonzero $\mathbb{Z}_2$ invariant and a pronounced intrinsic spin Hall response.

### B. Constructing Br kagome lattice on Ag(111) and STM characterization

The pristine electronic structure of freestanding Br kagome monolayer involves contributions from all three $p$ orbitals, including substantial out-of-plane $p_z$ character (Fig. S1 in SM [47]) [48–55]. To realize a $p_{x,y}$-orbital kagome system, Br was intentionally deposited on the Ag(111) surface, where $p_z$ orbitals are selectively suppressed through the orbital filtering effect [44], as shown in Fig. 2(a). Using $SnBr_2$ as a precursor, which provide a stable and clean Br flux, we achieved a well-ordered Br adlayer. A coverage-dependent structural evolution is observed (Fig. S2 in SM [47]), akin to previous reports [56,57]. Interestingly, when the Br coverage reaches ~0.33 ML, an extended, atomically flat kagome lattice without detectable competing phases is formed, as revealed by the large-area STM imaging shown in Fig. 2(b).

X-ray photoelectron spectroscopy (XPS) measurements confirm the adsorption of Br on the Ag(111) surface and rule out detectable Sn contamination (Fig. S3 in SM [47]). Step-height line-profile analysis (Fig. S4(a) in SM [47]) yield a value of 2.3 Å, in line with that of the Ag(111) substrate, indicating that the Br adlayer is a monolayer with uniform thickness. The sharp fast Fourier transform (FFT) pattern in Fig. 2(c) exhibits clear $C_6$-symmetry, confirming the long-range kagome order. Atomically resolved STM images [Fig. 2(d)] reveal a lattice constant of ~4.3 Å for the Br kagome monolayer (see Fig. S4(b) for detailed line profile [47]). One lattice vector aligns parallel to the straight step edge of the Ag(111) surface, allowing us to index the Br kagome as a 3×3 superstructure relative to the underlying Ag(111) surface [58,59]. A representative region of kagome lattice highlighted in Fig. 2(e) matches our STM simulation [Fig. 2(f)] and the structural model (Fig. S5 in SM [47]). In this model, the NN Br-Br distance is 4.33 Å, in excellent agreement with the experimental value.

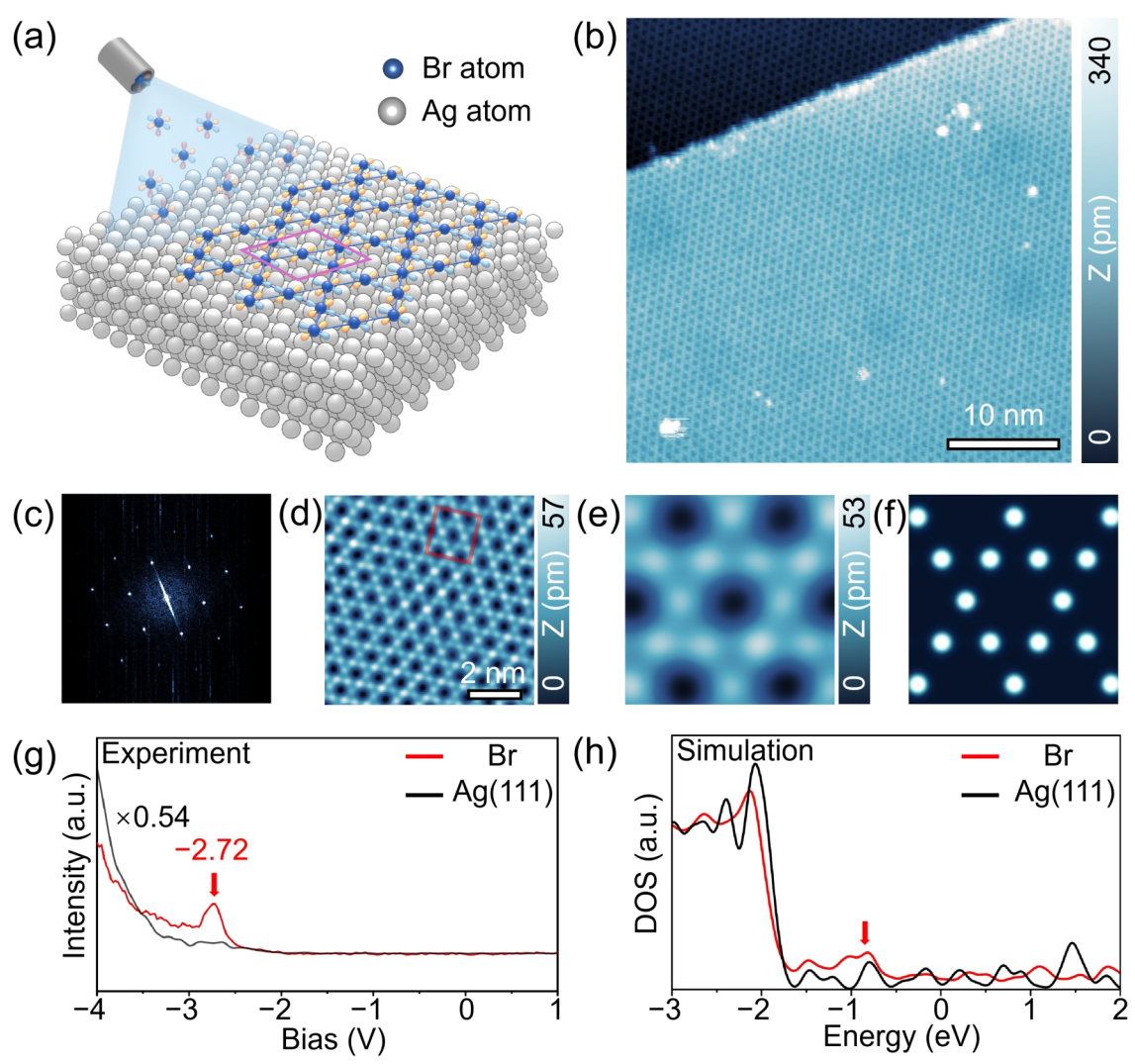

FIG. 2. STM characterization and local electronic structure of the Br kagome monolayer on Ag(111). (a) Schematics of $p_z$-orbital filtering when halogen atoms are adsorbed on the substrate. (b) Large-scale STM

image of the high-quality Br kagome monolayer, sample bias U= −2.00 V, tunneling current I= 80 pA. (c) FFT pattern of (b). (d) Atomically resolved STM image of the kagome lattice, U= 1.05 V, I= 120 pA. (e) A representative zoomed-in region marked by the red square in (d). (f) STM simulation of Br kagome monolayer on Ag(111). (g) *dI/dV* spectrum measured at the center of the kagome hexagon, together with a reference spectrum taken on pristine Ag(111) surface. The peak at −2.72 V marked by the red arrow is assigned to the Br *p* orbitals. (h) Calculated atom-resolved PDOS of the structure in (f), with the deep-energy feature dominated by the Br *p* orbitals highlighted by the red arrow.

To probe the electronic origin of this structure, we performed STS measurements. Figure 2(g) presents a differential tunneling conductance (*dI/dV*) spectrum measured at the center of the kagome hexagon, together with a reference spectrum taken on pristine Ag(111). A pronounced peak appears at ~−2.72 V exclusively on the Br-covered surface. By comparing this feature with the calculated atomic-orbital-resolved projected density of states (PDOS) [Fig. 2(h)], we reveal that it originates predominantly from the Br *p* states, while the Ag substrate contributes only a featureless background. Notably, the *dI/dV* spectra acquired at the kagome hollow site and atop the Br atoms show minimal spatial variation. This weak contrast indicates that the corresponding electronic states are not site-localized but are instead delocalized across the kagome network, consistent with the itinerant band picture derived from our tight-binding model.

### C. Electronic Structure Revealed by ARPES

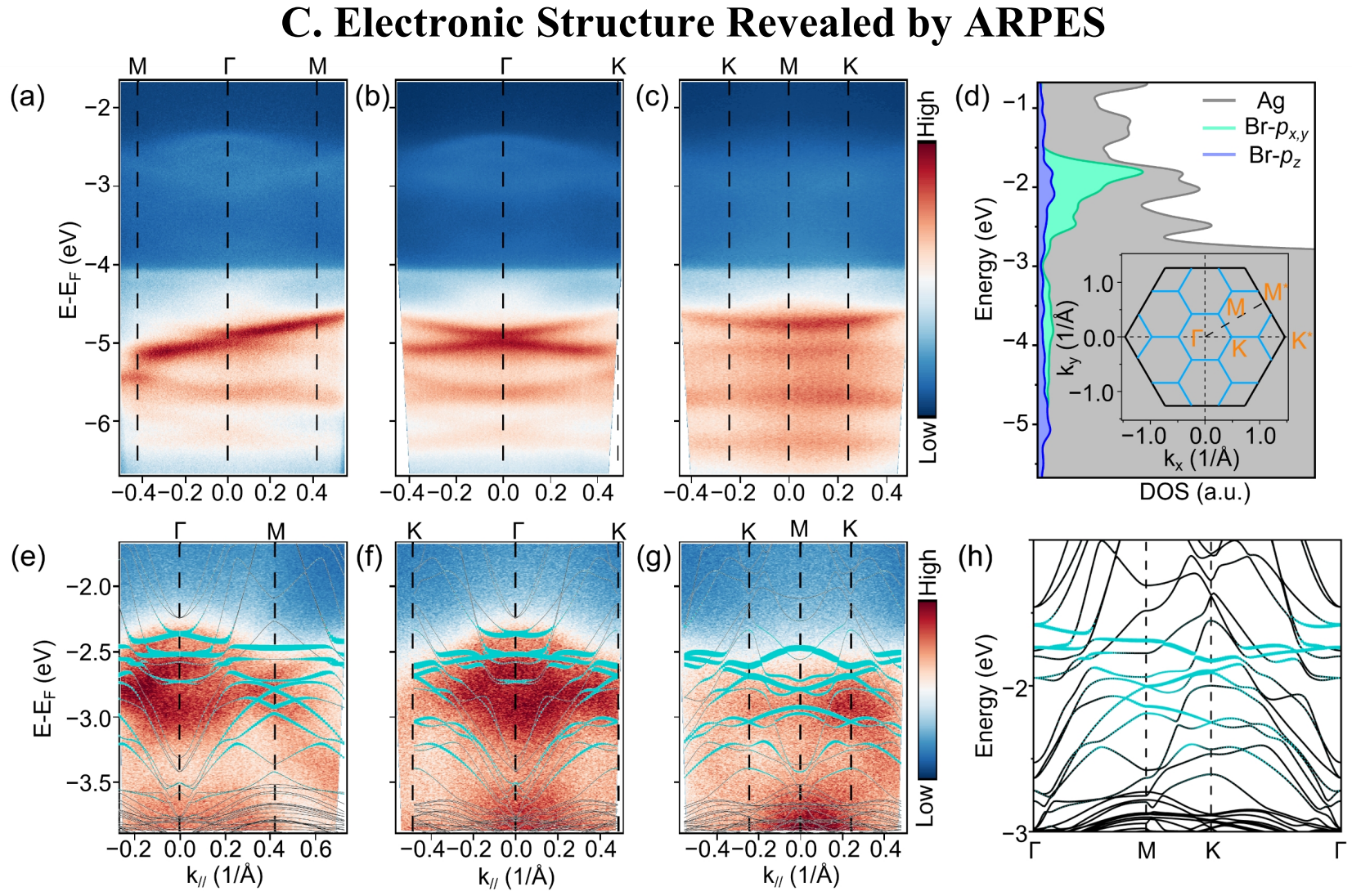

FIG. 3. Electronic band structure of Br/Ag(111). (a)-(c) ARPES intensity plots along the Γ-M, Γ-K and M-K directions of the Br BZ, respectively. (d) DFT-calculated PDOS. The inset compares the Ag (black) and Br (blue) surface BZ. (e)-(g) ARPES intensity plots in a zoom-in energy range to highlight the Br 4*p*-derived bands. The simulated Br band structure is overlaid on top. (h) DFT-calculated Br-$p_{x,y}$ orbital-projected (blue color) band structure of the Br/Ag(111) system, highlighting the dominant in-plane orbital character of the experimentally observed kagome-related bands.

To resolve the momentum-resolved electronic structure, we performed ARPES measurements on the Br/Ag(111) surface. The spectra in Figs. 3(a)-(c), plotted with respect to the surface Br Brillouin zone (BZ), reveal prominent band dispersions beginning approximately 4 eV below the Fermi level. These deeper features are primarily from Ag 4*d* orbitals, as corroborated by our density functional theory (DFT) calculations [Fig. 3(d)] and previous reports [60,61]. Their spectral weight also follows the Ag BZ periodicity (Fig. S6 in SM [47]). Within the [−2.5, −3.5] eV binding energy region, the bulk Ag *sp* band is not clearly observed because of the Br layer coverage. However, several bands exhibit a periodicity consistent with the Br BZ (Fig. S7 in SM [47]). The orbital-projected DOS of the Br/Ag(111) lattice identify these bands as predominantly Br 4*p* in orbital character [Figs. 3(d)], with the in-plane Br-$p_{x,y}$ states dominating the DOS, while the $p_z$ states are significantly broadened by hybridization with the Ag substrate.

A closer examination of the Br-derived bands is shown in Figs. 3(e)-(g). The observed dispersions deviate markedly from the textbook single-orbital kagome model, which features a Dirac cone at K, van Hove singularities at M, and an ideally flat band [62,63]. The Br-$p_{x,y}$ orbitals introduce bond-directional hopping that lifts the flat-band degeneracy and reshapes the dispersions near the high-symmetry points [31,64,65]. The calculated Br-$p_{x,y}$ projected band structure of Br/Ag(111) reproduces these modified dispersions with excellent fidelity [Fig. 3(h)]. This agreement between the measured ARPES spectra and the $p_{x,y}$-projected calculations confirms the successful isolation of the $p_{x,y}$ manifold via the orbital filtering mechanism, and establishes Br/Ag(111) as an experimental realization of the in-plane two-$p$-orbital kagome lattice.

### D. H-passivated effective model and topological properties

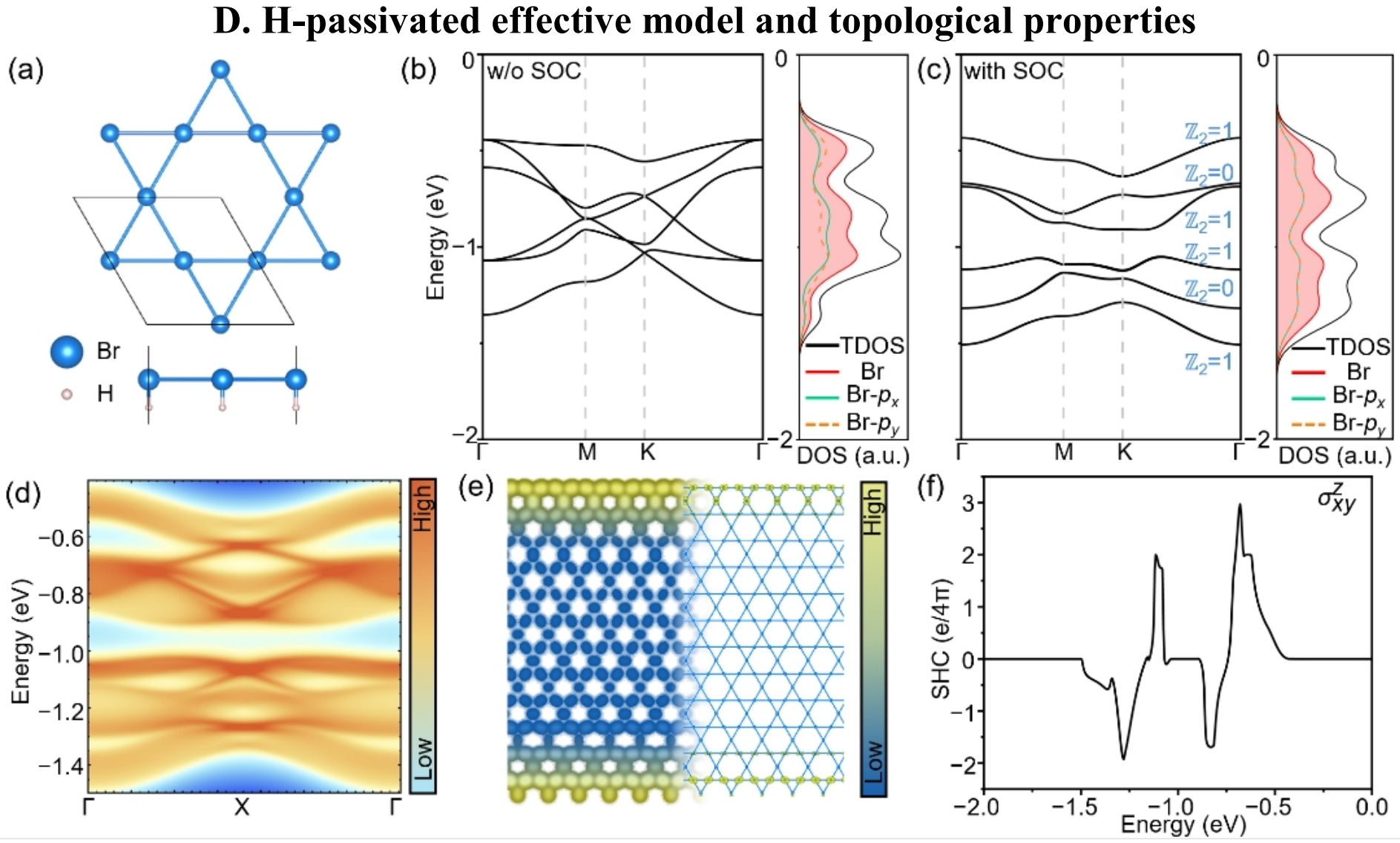

FIG. 4. Effective Br-$p_{x,y}$ kagome model and its topological transport properties. (a) Effective H-passivated Br monolayer kagome structure. (b,c) Orbital-projected band structures and PDOS without and with SOC, respectively; the $\mathbb{Z}_2$ indices obtained from the Wannier charge-center evolution are labeled in (c). (d) Surface spectral function along the Γ-X-Γ path, displaying in-gap topological edge states. (e) The STM simulation of the edge states at approximately −0.65 eV in (d). (f) Intrinsic chemical-dependent spin Hall conductivity $\sigma_{xy}^z$ in unit of e/4$\pi$.

To expose the intrinsic $p_{x,y}$-orbital kagome physics without the obscuring background from the Ag substrate, we construct an effective H-passivated Br monolayer kagome structure model [Fig. 4(a)] that retains the in-plane lattice constant of the Br kagome layer while introducing H atoms to mimic the effective substrate interaction for $p_z$-orbital filtering. The calculated band structures [Figs. 4(b)] show excellent agreement with those of the full Br/Ag(111) system [Fig. 3(h)] except their relative location with respect to the Fermi level, validating this simplified description (Fig. S6 in SM [47]). The PDOS confirms that the relevant bands are dominated by Br $p_x$ and $p_y$ orbitals, while the $p_z$ states are shifted away from the energy window of interest through coupling to the H-passivated environment. Maximally localized Wannier fitting further extracts the $p_{x,y}$ orbital basis and the associated hopping parameters, providing a direct connection to the tight-binding model introduced earlier.

In the absence of SOC [Fig. 4(b)], the system exhibits multiple symmetry-protected band degeneracies at Γ, M, and K, as well as along the K-Γ path. These features are characteristic of the double-orbital kagome model. Upon including SOC [Fig. 4(c)], a sizable global gap opens despite the light-element nature of Br. This gap originates from the first-order on-site $L_zS_Z$ coupling within the $p_{x,y}$ subspace, distinct from the normally expected higher-order SOC term for single-orbital systems, which efficiently lifts the degeneracies. The SOC-induced gap openings are accompanied by strong spin Berry curvature, naturally giving rise to a pronounced intrinsic spin Hall response.

To characterize the topology, we track the evolution of the Wannier charge centers (WCC). The $\mathbb{Z}_2$ invariant, determined by the parity of the WCC-flow crossings, yields the sequence of 1, 0, 1, 1, 0, 1 for the six bands from low to high energies, suggesting multiple topologically nontrivial gaps (detailed in Fig. S8 [47]). This nontrivial topology is further corroborated by the surface spectral function [Fig. 4(d)], which reveals topological edge states traversing the projected bulk gaps, particularly near the X point. The simulated STM image of a nanoribbon in Fig. 4(e), calculated at approximately −0.65 eV, visualizes the spatial distribution of these edge states (see Fig. S9 in SM for details [47]). Finally, the intrinsic spin Hall conductivity $\sigma_{xy}^{z}$, obtained from the Kubo formula within Wannier interpolation, exhibits pronounced energy-dependent peak structures [Fig. 4(f)]. These peaks arise from SOC-induced avoided crossings and the associated enhancement of spin Berry curvature, highlighting the strong spin-transport response intrinsic to the $p_{x,y}$ kagome bands.

## III. CONCLUSIONS

In summary, we have experimentally demonstrated a Br-$p_{x,y}$ kagome monolayer on Ag(111) as a viable platform for multi-orbital kagome physics. Combining STM/STS, ARPES, and first-principles calculations, we confirm an ordered kagome overlayer with Br-derived electronic states dominated by in-plane $p_{x,y}$ orbitals. In contrast to the conventional single-orbital kagome model, the present system hosts orbital-selective hopping and interorbital hybridization, which give rise to a characteristic $p_{x,y}$-orbital kagome band structure. An effective H-passivated model reproduces the measured band dispersions and captures the essential $p_{x,y}$ kagome band topology. Within this framework, SOC opens multiple global gaps and stabilizes a topologically nontrivial phase, characterized by a nonzero $\mathbb{Z}_2$ invariant, robust in-gap edge states, and an intrinsic spin Hall response. Although the metallic Ag substrate screens the kagome spectral weight and hinders direct observation of the predicted edge states, our findings motivate future measurements on weakly interacting or wide-bandgap insulating substrates. Our results extend kagome physics from the conventional single-orbital paradigm to an experimentally realizable multi-orbital regime, providing a feasible route toward orbital-engineered topology and spin transport in artificial kagome materials.

## ACKNOWLEDGMENTS

This work was supported by the National Key Research and Development Program of China (2022YFA1403500 and 2024YFA1408400), the National Natural Science Foundation of China (12274029, 12547158, 12204037, 12674069 and 12474474), and the Beijing Institute of Technology Research Fund Program for Young Scholars.